# Twitter-based traffic information system based on vector representations for words

**Sina Dabiri** [a,b*], **Kevin Heaslip** [a]

[a] Department of Civil and Environmental Engineering, Virginia Tech, Blacksburg, VA, USA
[b] Department of Computer Science, Virginia Tech, Blacksburg, VA, USA
[*] Corresponding Author
Email Addresses: sina@vt.edu, kheaslip@vt.edu

**Abstract**

Recently, researchers have shown an increased interest in harnessing Twitter data for dynamic monitoring of traffic conditions. Bag-of-words representation is a common method in literature for tweet modeling and retrieving traffic information, yet it suffers from the curse of dimensionality and sparsity. To address these issues, our specific objective is to propose a simple and robust framework on the top of word embedding for distinguishing traffic-related tweets against non-traffic-related ones. In our proposed model, a tweet is classified as traffic-related if semantic similarity between its words and a small set of traffic keywords exceeds a threshold value. Semantic similarity between words is captured by means of word-embedding models, which is an unsupervised learning tool. The proposed model is as simple as having only one trainable parameter. The model takes advantage of outstanding merits, which are demonstrated through several evaluation steps. The state-of-the-art test accuracy for our proposed model is 95.9%.

# 1. Introduction

In the past decade, social media networks have received much attention among ordinary people, agencies, and research scholars. Twitter is one of the fastest-growing social media tools that enables users to post and read short messages, called *tweets*. By means of Twitter applications on smartphones, users are able to immediately reports events happening around them on a real-time basis. The information disseminated by millions of active users everyday generates a new version of dynamic database that contains information about various topics. Since every out-of-home activity relies on transportation systems, people as dynamic social sensors might immediately report traffic conditions (e.g., slow traffic) and traffic incidents (e.g., accidents) by posting tweets while they are driving or observing the occurrence of a traffic-related event. Accordingly, Twitter data can be considered as a complementary traffic data source in addition to traditional sensors such as CCTV cameras and inductive loop detectors. However, unlike the fixed-point sensors, such data are ubiquitous and not limited by sparse coverage (Dabiri and Heaslip, 2018b). Harnessing this rich flow of user-generated contents provides inexpensive and widespread information on both traffic recurring congestion (i.e., predictable conditions such as daily rush hours) and non-recurring congestion (i.e., unpredictable incidents such as crashes). In particular, the specific objective of this study is to extract the traffic-related tweets that not only detect traffic incidents but glean the real-time level of serviceability of transportation systems. Such valuable information is beneficial to both drivers, in terms of finding shortest routes, as well as traffic managers, in terms of promptly recognizing unexpected behavior in traffic conditions.

Although a tweet object contains several attributes (e.g., id and created_at), several studies has focused on analyzing tweet's text to extract traffic information using machine learning algorithms. Machine learning techniques typically needs a fixed-length-numerical feature vector as input, which calls for an appropriate natural language processing method to model the text. Bag-of-words (BOW) representation is the most utilized approach for text representation. In this method, after transforming the tweet's text into a set of words, a type of occurrence measurement is assigned to each word as the feature value. Examples of feature values are binary occurrence of a word or a function of word-occurrence frequency (e.g., the classical term frequency-inverse document frequency (*tf-idf*)).

However, the BOW representation is subjected to some shortcomings. First, it only concerns the presence of words while the temporal order of words is discarded. Secondly, as the vocabulary size in the tweet corpus is very large and only a small subset of words is used in each tweet, the BOW matrix suffers from sparsity with many features values as zero and the curse of dimensionality. Notwithstanding that the curse of dimensionality can be addressed by keeping only high-important features using statistical feature selection methods, such techniques are at the risk of discarding essential traffic keywords. So some research has suggested modeling tweets on the top of only traffic keywords that have already been collected as the high-frequent words in traffic-related tweets. In spite of a dramatic dimension reduction by only taking a pre-defined set of keywords into account, the immediate criticism to using this concept is that the vocabulary may not include all important traffic-related keywords and is subjected to alter over time. New users, who their samples of tweets are not in the training set, might use different words to describe traffic conditions. Also, the twitter language contains many informal, irregular, and abbreviated words as well as a large number of spelling and grammatical errors (Atefeh and Khreich, 2015). Thus, representing tweets on the top of a pre-defined set of words (i.e., features) might not build a sophisticated classifier that is robust against tweet diversity from various datasets.

To address the above-mentioned shortcomings, in this paper, we develop an efficient and simple architecture on the top of word-embedding tools so as to classify tweets into non-traffic and traffic-related tweets. Word-embedding techniques map millions of words and phrases into numerical vector space in such a way that semantic-and-syntactic similar words are closer to each other. Word-embedding is a type of deep learning architectures that have recently been attracting a lot of interest in transportation fields (Dabiri and Heaslip, 2018a, Dabiri and Heaslip, 2019). We first identify the most frequent words in a small set of traffic-related tweets, called traffic keywords. Afterward, we measure the similarity between words in the subject tweet and traffic keywords through vector representation of words that have been obtained by word-embedding tools. A tweet that the average similarity of its words with traffic keywords exceed a trained threshold is classified as traffic-related. Training threshold, as the only parameter in our classification model, follows a very simple strategy without making any type of assumption. Through several evaluation steps, we demonstrate that the model performance is independent of a strict set of keywords, yet it only requires a small set of words related to traffic. In particular, unlike other studies, we show that traffic-related tweets without containing any traffic keyword are correctly inferred.

As one of the contributions in this study, we also collect and manually labeled 51,100 tweets using Twitter standard search application programming interface (API). To the best of our knowledge, a majority of studies have collected less than 5,000 labeled tweets for developing their twitter-based traffic information model while the maximum number of labeled tweets has been 22,000 tweets in (Gu et al., 2016). Since our proposed model does not require a large training set, such a massive labeled dataset gives room to show the model's superiority while distinct training sets are deployed.

This paper begins by related works in Section 2. After elaborating the process of collecting and labeling data in Section 3, it will then go on to describing the proposed framework in Section 4. We report findings and evaluate our proposed architecture through several steps in Section 5. Finally, we conclude the paper in Section 6.

**2. Related works**

The traditional automatic algorithms for detecting traffic events are divided into two groups based on the deployed traffic data acquisition system (Parkany and Xie, 2005): 1) fixed-point traffic flow sensors such as inductive loop detectors and video image processors, 2) probe vehicles that record vehicles' spatiotemporal information of when they are performing their regular trips using positioning tools such as GPS. Afterward, mathematical algorithms (e.g., pattern matching, statistical discriminative algorithms, and time series methods (Gu et al., 2016)) are employed to capture the unexpected behavior in traffic flow characteristics. Comprehensive and systematic literature reviews on various types of traditional traffic-event-and-incident-detection algorithms are available in the survey papers (Williams and Guin, 2007, Parkany and Xie, 2005). Beyond the traditional traffic data sources, the growth in popular social networking tools has been generating a massive data source with a high coverage area that contains rich information about transportation infrastructures and travel behavior. Up to now, several studies have analyzed such rich and inexpensive data to support decision making in a variety of transportation-related applications, including travel recommendation systems, travel demand analysis, travel patterns and human mobility, urban planning, incident detection, and emergency systems (Rashidi et al., 2017). However, in order to be alongside with the main focus of this study, we only review those works that have proposed a framework for extracting traffic-related information from Twitter data.

Although some studies have used non-text attributes of tweets for their analysis (Chaniotakis and Antoniou, 2015), the tweet text has received more attention for monitoring traffic conditions and detecting traffic incidents. These studies are categorized into two groups depending on how a tweet's text is represented: 1) Non-numerical features. In this group, tweets are represented based on whether or not a tweet contains a pre-defined set of keywords. 2) Numerical features. Unlike the first group, tweets are mapped to numerical feature vectors, typically using BOW representation. Afterward, a discriminative method has been utilized to infer traffic-related and/or traffic-incident tweets from a stream of tweets.

One of the earliest work from the first category was conducted by (Wanichayapong et al., 2011). The key goal in their proposed model was to differentiate between traffic-point and traffic-link tweets. The former associates with only one point (e.g., a crash) while the latter type associates with a road-start point and a road-end point (e.g., traffic jam). A tweet is tokenized and parsed into four word categories: 1) places (e.g., roads), 2) traffic problems (e.g., accident), 3) words indicating start and end locations of traffic events, 4) ban words (i.e., vulgarity/profanity words). Since tweets are not represented with numerical features, rule-based and heuristic methods are applied to first filter traffic-related tweets and then classify traffic tweets into point and link types. All heuristics are adopted based on what types of word categories a tweet contain. Another example of the first category is the work by (Rebelo et al., 2015), who utilized different variants of Pearson Correlation Coefficients to measure the correlation of traffic volumes and locations between desired tweets and ground-truth tweets. The selected tweets are those that contain traffic-related keywords. (Ribeiro Jr et al., 2012) developed their Traffic Observatory framework to detect traffic conditions and traffic events using Twitter data, geocode them, and display them on the Web. Traffic-related tweets were manually identified by searching a static and predefined list of words (e.g., intense, slow, regular, and free) that express traffic situations. However, the focus of the study was to locate traffic-related tweets by performing both exact and fuzzy matching using an enhanced gazetteer that contains urban details.

With regard to the numerical features category, (Carvalho, 2010) was one of the earliest studies that used BOW to model tweets. Support Vector Machine (SVM) was deployed to infer traffic-related tweets with either uni-grams or bi-grams as features. In a more systematic analysis by (D'Andrea et al., 2015), after applying some preprocessing steps, the tweet corpus was turned into numerical feature vectors using BOW represenation. Next, the most relevant features were selected by computing Information Gain (IG) for all features. Features with positive IG were selected and fed into the SVM algortihm to be classified into one of the three groups: non-traffic, traffic due to congestion, and traffic due to external events. In a study conducted by (Fu et al., 2015), first, traffic-incident-related words with the highest *tf-idf* weights were extracted from tweets posted by some influential user accounts. Using the Apriori algorithm, the most frequent word set was identified to build more efficient queries for crawling traffic-related tweets through Twitter APIs. Finally, the extracted tweets were scored and ranked based on the sum of traffic keywords *tf-idf* weights. In a follow-up and more comprehensive study by (Gu et al., 2016), an adaptive data acquisition process for building a dictionary with traffic-related words was suggested. The words in the built dictionary were served as features in the Semi-Naïve-Bayes (SNB) model to classify tweets into traffic incident (TI) versus non-traffic incident groups. A recent study by (Pereira et al., 2017) integrated the BOW and word embedding to create a group of features for each tweet.

Using the created feature vectors, classical supervised learning algoritihms (e.g., SVM) were deployed to identify travel-related tweets from non-related ones.

Except the work by (Pereira et al., 2017) that utilized word embedding in addition to BOW, no study has constructed their model purely based on distributed vector representations of words. In terms of classification, almost all studies have leveraged traditional supervised learning algorithms. However, in this study, we propose a simple architecture that is primarily contingent on word-embedding techniques, which is an unsupervised learning algorithm. Furthermore, our classification includes only one training parameter, which in turn requires a low sample of tweets for training.

## 3. Data collection procedure

We have collected 51,100 tweets using the free-of-charge standard search API. Originally, we have manually labeled them into three classes, which are defined as follows.
  (1) *Non-Traffic (NT)*: Any tweet that does not fall into the other two categories is labeled as NT.
  (2) *Traffic Incident (TI)*: This type of tweet reports non-recurring events that generate an abnormal increase in traffic demand or reduces transportation infrastructure capacity. The examples of non-recurring events include traffic crashes, disabled vehicles, highway maintenance, work zones, road closure, vehicle fire, traffic signal problems, special events, and abandoned vehicles. Since the ultimate goal of our framework is to inform users and agencies on the occurrence of a traffic incident in a real-time basis, if a tweet reports on the clearance or re-opening of roads that had already been affected by non-recurring traffic events, that tweet is classified as TCI, the third tweet category. Indeed, such tweets are providing information on the current status of the network rather than informing an ongoing traffic incident.
  (3) *Traffic Conditions and Information (TCI)*: This type of tweet reports traffic flow conditions such as daily rush hours, traffic congestion, traffic delays due to high traffic volume, and jammed traffic. Also, any tweets that disseminate new traffic rules, traffic advisory, and any other information on transport infrastructures (e.g., new facilities or changing the direction of a street) are classified as TCI.

As we focus on a binary classification in this study, we build only a 2-class dataset in which tweets are categorized into traffic-related tweets (i.e., TI and TCI), denoted as (TR) and non-related-traffic tweets (NT).

In order to have an efficient strategy for collecting and labeling tweets, we first build a traffic-related dictionary through Twitter's profiles who mainly disseminate traffic information, also called Influential Users (IUs) in literature. Using the built traffic dictionary and IUs, we have collected tweets in three datasets with different procedures. We elaborate these procedures after describing how the dictionary is built.

*3.1. Traffic-related dictionary*

The traffic-related dictionary consists of only the most frequent words occurred in traffic-related tweets. A common method in literature for obtaining traffic-related tweets is to returning them from accounts that belong to transportation and emergency service providers (e.g., State Department of Transportation (DOT)). However, a chunk of tweets from such users are not related to traffic information and incidents. As a consequence, a time-consuming labeling process must be implemented to distinguish traffic-related tweets from non-traffic-related tweets. A more effective way to fetch traffic-related tweets with a very high chance is to provide a list of Twitter users that primarily post traffic information.

511 is a national traffic information telephone hotline across some regions of the United States that only provides traffic-related information for drivers. Thus, we create an initial list of Twitter users that are relevant to the 511 by making queries based on the keywords "511" and "Traffic" through the API GET users/search method. After obtaining the matched users, we simply visualize the recent tweets of the matched users to ensure that these Twitter accounts primarily disseminate traffic incident information. Following this process, we obtained a IU list with 69 users such as "511nyNJ, TN511, WV511South, fl511_northeast, and 511northwestva". A few examples of traffic-related tweets extracted from IUs are provided in Table 1.

**Table 1**
Examples of traffic-related tweets extracted from Influential Users (IUs)

| User screen name | Traffic-related tweets |
| --- | --- |
| 511northernva | Cleared: Disabled Vehicle: EB on I-66 at MM68 in Arlington Co.11:36 AM |
| 511Georgia | Accident, I-285 North past Lavista Road, far right lane blocked. #511GA |
| 511NYC | Closure on #ThrogsNeckBridge SB from Bronx side to Queens side |

Subsequently, we return a pool of the most recent tweets posted by the users in the IU list. Following the BOW concept, the collected traffic-related tweets are tokenized and the token-occurrence frequency is computed for each distinct token. Occurrence frequency of tokens are then summed up over all tweets. Finally, tokens with the highest occurrence frequency are selected to build the traffic-related dictionary. Since a portion of collected tweets might be non-traffic-related, we re-investigate the selected words one by one to make sure all words have a semantic relationship with traffic and transportation. Examples of traffic words in the dictionary are "traffic, blocked, lane, construction, crash, congestion, delays, vehicle, incident, ramp, and street".

Having the IU list and traffic-related dictionary in hand, we collect tweets in three datasets with different API methods. The rationale behind collecting tweets with different procedures is to only expedite the process of data collection since each dataset brings its own advantages. After manually labeling tweets in these datasets, they are concatenated to form one comprehensive-labeled-tweet dataset. Training and test sets are then randomly selected from this comprehensive-labeled-tweet dataset.

*3.2. Collecting the first dataset*

This dataset contains English tweets from random users located within the USA and the south part of Canada. Returning a traffic-related tweet (i.e., TI and TCI tweets) among all other types of tweets is similar to outlier detection. Thus, if no keyword is specified for fetching tweets from random users, the chance of receiving traffic-related becomes too low. For addressing this problem and making a balance between traffic-related and non-traffic related tweets, we utilize the words in the traffic-related dictionary for making queries.

So the query in the API GET search/tweets method is specified as a combination of words in the traffic-related dictionary. Such a query increases the chance of getting traffic-related tweets generated by random users. In each request, we use a combination of two, three, or four words from the built dictionary. We collected almost 16,000 tweets from random users. Since our goal is to obtain the same portion of tweets for both traffic-related and non-traffic related tweets from random users, we removed the query parameter after obtaining almost 8,000 traffic-related tweets. Fetching tweets without queries made by traffic-related keywords dramatically soars the chance of getting non-related traffic tweets.

*3.3. Collecting the second dataset*

In this step, we aim to collect only *potential* traffic-related (i.e., TI and TCI) tweets. Therefore, we need to retrieve tweets for this dataset from Twitter accounts in the IU list. As a consequence, there is no need for making queries based on some keywords. Instead, tweets are collected from IU users using the API GET statuses/user_timeline method. For this dataset, we collected almost 17,000 tweets from users in the IU list. It should be noted that although tweets in this set are potential to be traffic-related tweets, they have manually inspected and labeled into one of the NT, TI, or TCI category.

*3.4. Collecting the third dataset*

In this dataset, we aim to collect potential non-related traffic tweets. Analogous to the second dataset, in order to speed up the process of collecting and labeling tweets, we fetched tweets from users who almost never post tweets about traffic condition. We provide a list of 70 users from the fashion industry, politicians, celebrities, and irrelevant organizations to the traffic incident management. Examples of such users are "AshleyFurniture, AEO, nabp, MikePenceVP, and verizon". Again, using the API GET statuses/user_timeline method, we collected roughly 17,000 recent tweets from 70 non-traffic-related users. Note that although tweets in this set are potential to be non-traffic-related tweets, they have manually inspected and labeled into one of the NT, TI, or TCI category. Because the chance of retrieving traffic-related and non-traffic-related tweets are high in the second and third datasets, respectively, the process of labeling tweets for these two sets are faster compared to the first dataset.

We apply the same preprocessing steps as described in Methodology Section to tweets' texts so as to facilitate the labeling process. The label distribution of tweets in the three datasets is shown in Table 2. For having the same number of traffic-related tweets (i.e., TI and TCI) and non-traffic-related tweets (i.e., NT), we trim back some tweets. The last row in Table 2 shows the tweet distribution among three labels as the final labeled-tweet dataset after removing some tweets. We randomly choose 80% of labeled tweets (40,879 tweets) as the training set while holding out the remaining 20% (10,221) as the test set. Tweets from each class have the same proportion in training and test sets.

**Table 2**
Number of tweets in each dataset and class

| Dataset\Class | NT | Traffic-Related (TR) | | Total |
| --- | --- | --- | --- | --- |
| | | TI | TCI | |
| First | 8,138 | 5,414 | 2,651 | 16,203 |
| Second | 377 | 12,022 | 5,462 | 17,861 |
| Third | 18,259 | 2 | 0 | 18,261 |
| Total | 26,774 | 17,438 | 8,113 | 52,325 |
| Final dataset | 25,550 | 17,437 | 8,113 | 51,100 |

## 4. Methodology

The specific objective of our Twitter-based framework is to automatically classify tweets into two NT and TR groups, where TR collection constitutes both traffic incidents and traffic conditions. It is worth mentioning that the proposed method is applicable for any type of binary classification. Accordingly, for incident detection applications, one might consider only traffic-incident tweets against non-traffic ones. Three primary blocks constitute our Twitter-based traffic information system: 1) identifying traffic keywords, 2) distributed vector representations for words, 3) classification task.

It should be noted that the following pre-processing steps are applied to tweet text at any time a tweet is used in our analysis:
- Remove all links (web URLs) from the text.
- Remove all special characters and punctuation marks.
- Remove stop words (e.g., articles and prepositions)
- If the current tweet is the result of re-tweeting another tweet, the original text from the native tweet is used to avoid having a truncated text.
- Substitute all U.S. and Interstate Highways with the word 'highway'. 'US Number', 'I-Number', and other similar formats are considered as the patterns for finding U.S. and Interstate Highways in texts.
- Remove the remaining numbers.

Applying the above-mentioned preprocessing steps to the first tweet in Table 1, as an example, and then transforming it into a group of words results in: ['cleared', 'disabled', 'vehicle', 'eb', 'highway', 'MM68', 'arlington', 'am']

*4.1. Identifying traffic keywords*

The first step in our methodology is to provide a small set of traffic-related keywords. As long as the keywords are semantically related to traffic incidents and conditions, the number of keywords and the source of obtaining these keywords does not have a significant impact in the prediction quality of our model. This is a notable advantage of our proposed network, which will be demonstrated later in Section 5. Thus, various approaches can be utilized to define the list of traffic keywords. The simplest way is to manually generate traffic keywords such as: 'traffic, incident, blocked, crash, lane, street, left, caution, roadway, congestion, avenue, marker'. A more reliable approach is to extract the most frequent words in a pool of TR tweets, in which words with high-occurrence frequency are determined as traffic keywords. As will be shown in Section 5, there is no need for having a large size TR-tweet collection to obtain frequent keywords, which counts as another advantage of the framework.

*4.2. Distributed vector representations for words*

A TR tweet is differentiated from NT tweets based on how much its words are semantically similar to the traffic keywords collected in the previous step. Semantic similarity between words can be captured through a word embedding technique, which maps millions of words to vectors of real numbers in such a way that words with similar meanings tend to be closer to each other in vector space. To learn vector representations of words, we employ the word2vec model proposed in the seminal study by Mikolov et al. (Mikolov et al., 2013). Word2vec model is an unsupervised learning algorithm that deploys a Neural Network Language model (NNLM) to learn high-quality distributed representations of words in vector space from a massive data sets with millions of words in the vocabulary. In their continuous BOW (CBOW) architecture, the main task is to predict the occurrence probability of a word given other words in its context. The NNLM is a fully connected neural network with one hidden layer. The number of neurons in the input and output layers are equal to the number of words in the vocabulary, denoted as $V$. The hidden layer size is set to the dimensionality of the word vectors, denoted as $D$. Thus, the weight matrix **W** between the input and hidden layers has the size $V \times D$. However, the input to the network is encoded using 1-of-V coding which activates only the words in the surrounding window of the target word at any given time while the rest are set to zero. Thus, if we have a sequence of training words [$w_1$, $w_2$, …, $w_v$, …, $w_{V-1}$, $w_V$], the objective of the model is to maximize the average log probability of predicting a word $w_v$ given the words in its context window as shown in the equations (1) and (2):

$$(maximize)_v \quad \frac{1}{V} \sum_{v=k}^{V-k} \log p(w_v | w_{v-k}, \dots, w_{v+k}) \tag{1}$$

$$p(w_v | w_{v-k}, \dots, w_{v+k}) = \frac{e^{y_{w_v}}}{\sum_{v \epsilon V} e^{y_v}} \tag{2}$$

where $k$ is the context window with a typical value of 5. In Equation (2), the softmax function is utilized to generate a probability distribution over all the words in the vocabulary, where $y_v$ denotes the activation vector for the word $v$. Since the cost of computing Equation (2) is too much in practice, the full softmax is approximated by the hierarchical softmax with a binary tree representation of the output layer (Mikolov et al., 2013). The NNLM model is trained using stochastic gradient descent via backpropagation. The rows in the matrix **W** (i.e., the word vectors) are the final outputs of the word2vec framework. In parallel to the CBOW, the continuous skip-gram architecture was also introduced. While the CBOW architecture predicts the current word based on its context window, the skip-gram predicts surrounding words given the current word (Mikolov et al., 2013). The CBOW and continuous skip-gram architectures can be implemented in a huge corpus of tweets to produce the new vector representation of words using the available online word2vec tool (Google, 2013).

In this study, we use two pre-trained word2vec models that are publicly available: 1) Google word2vec that has been trained on the part of the Google News dataset with almost 100 billion words. Its weight matrix **W** contains 300-dimensional vectors for 3 million words and phrases (Google, 2013), 2) Twitter word2vec that has been trained on 400 million tweets. Its weight matrix **W** contains 400-dimensional vectors for almost 3 million words and phrases (Godin, 2015).

*4.3. Classification Task*

Figure 1 represents the steps for classifying unlabeled tweets into NT and TR groups. The core idea in the classification task is to determine the extent that a tweet is similar to the identified traffic keywords in the first step. For achieving this goal, first, the vector representations of traffic keywords are averaged and stored into the vector **MU**. This vector is used as the vector representation of all words in traffic keywords. Afterward, we clean the target tweet with text-processing steps and then tokenize it into a group of words, denotes as **S**. Then, for each word $v$ in **S**, we compute the cosine similarity between the vector representation of $v$, denoted as $\mathbf{H}_v$, and the vector **MU**. Next, we obtain the average of cosine similarity for all words in **S** as follows, called Tweet Similarity Index (*TSI*).

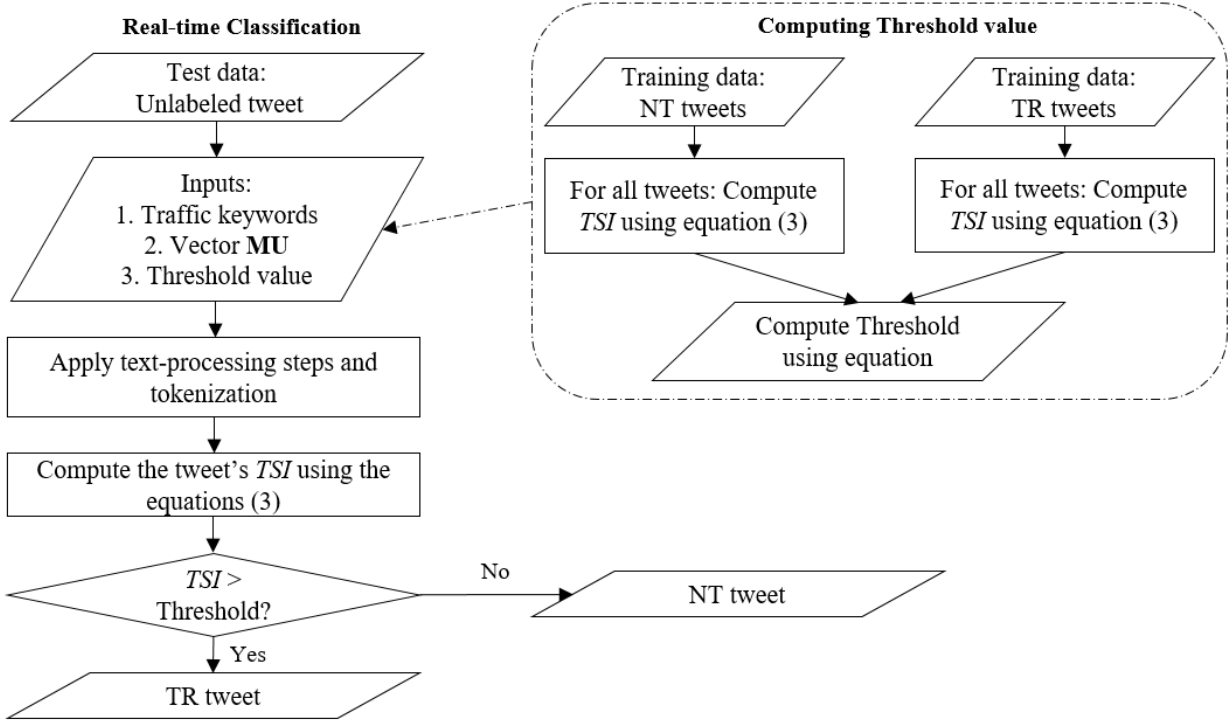

Fig. 1 Flowchart for classifying tweets into NT and TR groups.

$$TSI = \frac{1}{|\mathbf{S}|} \sum_{v \in \mathbf{S}} \frac{\mathbf{MU} \cdot \mathbf{H}_v}{\|\mathbf{MU}\|_2 \cdot \|\mathbf{H}_v\|_2} \tag{3}$$

A tweet is classified as TR if its computed *TSI* exceeds a threshold value. *Threshold* is the only parameter in our framework that essentially needs the labeled data to be determined. Denoting the average of *TSI* for all labeled NT and TR tweets as $NT_{TSI}$ and $TR_{TSI}$, respectively, *Threshold* is computed using Equation (5):

$$Threshold = \frac{NT_{TSI} + TR_{TSI}}{2} \tag{4}$$

Similar to traffic keywords, learning *Threshold* value does not require a large training set, which can be considered as another advantage of the model.

As mentioned, our model is as simple as possible without requiring to make any assumption. For example, we do not hypothesize that the vectors of words in the traffic keywords have drawn from a reference probability distribution (e.g., multivariate normal distribution). Analogously, we do not assign any distribution to *TSI* for NT and TR tweets and then predict based on the captured distribution. Another outstanding feature of the prediction model is that even if a target tweet has no words from the traffic keywords, it is not necessarily classified as the NT category. However, methods that model tweets using BOW representation on the top of traffic keywords definitely classify a target tweet as NT if it does not contain words from traffic keywords (Gu et al., 2016). In our methodology, the primary criterion for classifying a tweet is the tweet's semantic relationship with a small set of traffic keywords rather than the existing of traffic keywords in the target tweet. Moreover, unlike BOW representation, the proposed method does not suffer from the curse of dimensionality and sparsity. This also prevents discarding some words on account of reducing dimensionality by means of statistical feature selection models (D'Andrea et al., 2015) or pre-defined sets of traffic keywords (Fu et al., 2015, Gu et al., 2016). Finally, as will be demonstrated in the following section, the proposed model does not need a large training set for neither extracting traffic keywords nor training the *Threshold* parameter.

## 5. Results and discussion

All analyses have been coded in the Python programming language. In particular, we have used two Python-based libraries, scikit-learn and Gensim, for processing and extracting features from tweets. We assess the effectiveness of our model in three sorts of evaluation: 1) Quantitative assessment, 2) Comparison with other methods, 3) Qualitative assessment.

*5.1. Quantitative assessment*

As we mentioned earlier, neither threshold nor traffic keywords need a large training set. Accordingly, for all experiment steps in this section, we randomly sample one 1000-tweet training set from the main training dataset with 40,879 tweets to extract traffic keywords and then compute the only parameter *Threshold* using another distinct 1000-tweet training set.

First of all, we intend to show how the proposed model works. We extract some traffic keywords from a 1000-tweet training set. Then, using extracted keywords and another 1000-training set, we compute *TSI* for all NT and TR tweets based on Equation 3. Figure 2 illustrates and compares the histograms of *TSI* values for NT and TR tweets. As can be seen, both histograms are almost symmetric and unimodal where their centers are quite apart from each other. The similarity in the variance of the NT and TR distributions, which are equivalent to 0.04 for both distribution, is convincing evidence for considering the mean value as the discerning characteristic of the two distributions. The vertical red and blue lines show the average of *TSI* for NT ($NT_{TSI}$) and TR ($TR_{TSI}$) tweets, respectively. Based on Equation 4, *Threshold* as the only parameter in the model is determined as the average of $NT_{TSI}$ and $TR_{TSI}$ as shown with the vertical black line. *Threshold* separates the two distributions quite well from each other although a small portion of each set is misclassified.

In the second step, we inspect how the prediction quality of our model changes by varying the number of traffic keywords. We use accuracy on the 10,221-tweet test dataset as the performance measure. According to Figure 3, the prediction quality sharply increases to roughly 95% while having only 8 traffic keywords. The test accuracy remains constant around 95-96 % for the number of keywords between 8 to 100. This demonstrates the proposed methodology is almost insensitive to the number of traffic keywords as long as we have around 10 traffic keywords.

In the third step, we demonstrate the model robustness by reporting the test accuracy on multiple training sets. Using the main training dataset with 40,879 tweets, we construct 39 separate training sets that are randomly sampled from the main training set. Each of these training sets constitutes two distinct 1000-tweet sets, one for extracting traffic keywords and one for computing the *Threshold* value. It should be noted all 39 training sets are completely separated. For every training set, we compute *Threshold* using both Twitter and Google word2vec models. The built models are then tested on the same test dataset. Figure 4 shows the test accuracy for various training sets based on both Twitter and Google word2vec models. It is apparent from Figure 4 that our methodology achieves the same test accuracy with less than 1% variation while trained on separated training sets with two types of word2vec models. Another important finding is that the model obtains a high and similar rate of accuracy as long as utilizing well-trained word2vec models such as pre-trained Google and Twitter wod2vec models. The state-of-the-art test accuracy for our proposed model is 95.9%.

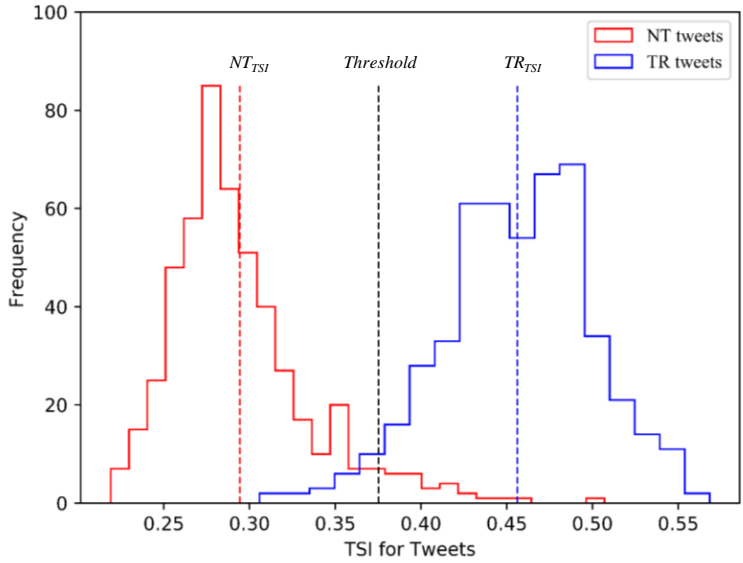

Fig. 2 Histograms of *TSI* values for NT and TR tweets in the training set. The red, black, and blue vertical lines show the *NT$_{TSI}$*, *Threshold*, and *TR$_{TSI}$* values, respectively.

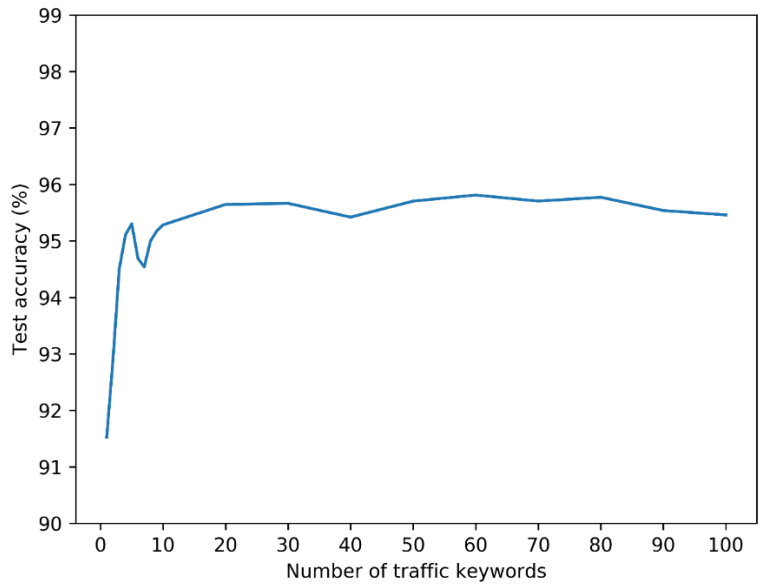

Fig. 3 Examining the test accuracy of the proposed model for varying number of traffic keywords.

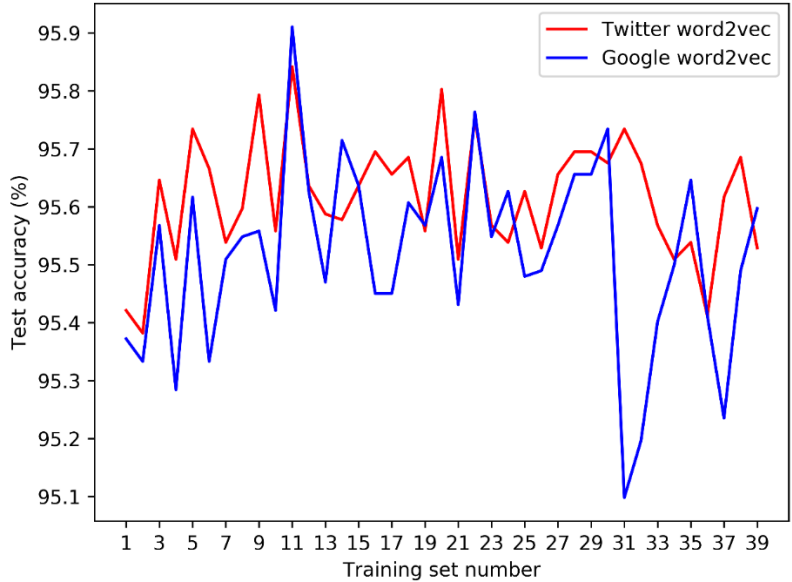

Fig. 4 Test accuracy of the proposed model on multiple separated training sets, using Twitter (red) and Google (blue) word2vec models.

Although the traffic keywords extracted from these 40 training sets are different, there are a high degree of overlaps between their first 10 keywords. The 10 most-frequent-traffic keywords for almost all training sets include: 'highway', 'exit', 'lane', 'crash', 'blocked', 'cleared', 'closed', 'updated', 'right', 'traffic'. So in our fourth step of evaluation, in order to illustrate how the accuracy might change if different sets of traffic keywords are fed into the framework, we utilize the first, second, third, and fourth 10-most-frequent traffic keywords in a random 1000-tweet training set, as shown in Table 3. These sets are selected in order from the first 50 most-frequent keywords. In addition, we add another set in the last column, which contains non-traffic keywords. Again, the results confirm the high quality of model as long as we have a set of traffic keywords, regardless of their frequency rank. This shows even if the less frequent yet traffic-related keywords are utilized, the algorithm performs well. Nonetheless, the model has a poor prediction with 30% accuracy when non-traffic keywords are used, which highlights the importance of traffic keywords.

Achieving high accuracy is only a positive starting point for having a reliable classifier. So in the final step, we assess the model by reporting other evaluation measures such as confusion matrix, recall, and precision, as shown in Table 4. What obviously stands out in Table 4 is high precision, recall, and F-score for both labels. In particular, the high precision results in the very low false-alarm rates, which is one of the most negative points in the traditional AID algorithms (Williams and Guin, 2007). Furthermore, due to the same values of precision and recall, we conclude that the model misclassifies both NT and TR at the same rate.

**Table 3**
Test accuracy of the proposed model trained with various set of keywords

| # most frequent | Keywords | Test accuracy (%) |
|---|---|---|
| First | 'highway', 'exit', 'lane', 'updated', 'lanes', 'cleared', 'blocked', 'closed', 'crash', 'traffic' | 95.3 |
| Second | 'station', 'right', 'sb', 'north', 'vehicle', 'rd', 'accident', 'construction', 'st', 'south' | 94.5 |
| Third | 'nb', 'left', 'incident', 'street', 'eb', 'sr', 'ramp', 'directions', 'turnpike', 'new' | 94.8 |
| Fourth | 'mm', 'ave', 'mile', 'update', 'pa', 'delays', 'wb', 'southbound', 'delay', 'west' | 95.8 |
| Non-traffic | 'milk', 'book', 'mouse', 'tea', 'computer', 'shoes', 'watch', 'pen', 'chair', 'tissue' | 30.4 |

**Table 4**
Confusion matrix, recall, and precision for the proposed model

| **Proposed model** | | Predicted Class | | |
|---|---|---|---|---|
| | | NT tweet (count) | TR tweet (count) | Recall% |
| Actual Class | NT tweet (count) | 4998 | 212 | 95.9 |
| | TR tweet (count) | 206 | 4905 | 96.0 |
| | Precision% | 96.0 | 95.9 | Average F-score: 95.9% |

*5.2. Comparison with other methods in literature*

In this section, we compare the performance of our framework with the most relevant studies in literature. Since our methodology utilizes numerical feature vectors for tweet representation, we only reproduce the works in the second group introduced in Section 2: Numerical features. As the codes corresponding to the frameworks of the relevant studies are not publically available, we do our best to replicate their work. During the implementation, we assume a reasonable setting for any case that the required information has not been provided. Finally, we train and test their work on our training and test datasets to have a fair comparison.

Table 5 compares the test accuracy of our methodology with several relevant frameworks published in literature. The methods for tweet represntation and classification are also provided in Table 5. Our simple yet efficient model is competitive with state-of-the-art ones in literature. Among models in literature, achieving the highest accuracy by the work in (Pereira et al., 2017) confirms the effectiveness of word embedding. Nonetheless, their proposed model is not only built upon on BOW with curse of dimensionality and sparsity, but also deprived of capabilities in our proposed model that were demonstrated in the previous section.

*5.3. Qualititave Assessment*

Since the core idea for the classification task is to obtain the similarity between words in the target tweet and traffic keywords, it could be interesting to extract the most similar words to some of the most-frequent words in the whole training dataset from both word2vec models. Table 6 represents the most similar words in Twitter and Google word2vec models to a few examples of highly-frequent-traffic keywords including 'highway', 'exit', 'traffic', 'crash', and 'vehicle'. Note that we discard similar words that have the same root or stem as their corresponding traffic keywords.

**Table 5**
Performance comparison between our model and relevant studies in literature

| Framework similar to: | Tweet representation | Classification | Accuracy (%) |
|---|---|---|---|
| Carvalho (2010) | BOW based on basuni-gram or bi-gram | SVM | 94.9 |
| D'Andrea et al. (2015) | BOW based on feature selection based on IG | SVM | 91.8 |
| Fu et al. (2015) | Apriori algorithm & frequent keywords | Ranking | 65.1 |
| Gu et al. (2016) | BOW based on traffic keywords | SNB | 93.0 |
| Pereira et al. (2017) | Combination of BOW & word embedding | SVM | 95.4 |
| This study | Word2vec models | Measuring TSI | 95.9 |

**Table 6**
Most similar words in Twitter and Google word2vec models to examples of highly-frequent-traffic keywords

| Highly-frequent-traffic keywords | Most similar words | |
|---|---|---|
| | Twitter word2vec | Google word2vec |
| highway | freeway,interstate,motorway,expressway | expressway,roads,roadway, motorway |
| exit | ramp,entrance,onramp,collector-distributor | departure,entrance,withdrawal, egress |
| traffic | rush-hour,bumper-to-bumper,congestion, motorway | congestion,flagmen,wreck-snarls, ticketing-speeders |
| crash | pileup,collision,accident,wreck | accident,collision,wreck,rollover |
| vehicle | car,truck,tractor-trailer,watercraft | car,SUV,minivan,jeep |

**Table 7**
Examples misclassified tweets with their actual and predicted labels

| # | Tweet texts | Actual label | Predicted label |
|---|---|---|---|
| 1 | Is there a street or sidewalk left in Cambridge that is not under construction? | NT | TR |
| 2 | There's heavy police presence in the area of Military Avenue and Leo Street in Green Bay. | NT | TR |
| 3 | The incident happened at the Edwin Hotel, which is under construction at Walnut Street. | NT | TR |
| 4 | Congestion: Dolphin Exwy EB - Between NW 107th Av and NW 87th Av/Galloway Rd - slow traffic | TR | NT |

Notwithstanding that there is somewhat overlap between founded words in Twitter and Google word2vec models, similar words achieved by Twitter are more meaningful and useful in terms of traffic concepts. A possible explanation for this fact is that Twitter model has been trained on a tweet corpus, which is more related to the task-at-hand.

We are also curious to investigate for what type of tweets our model is unable to correctly predict their label. Table 7 presenets some examples of missclassfication in the test set. Regarding to false-positive predictions, there are some ambuguities in the tweets that might even mislead humans. For example, the tweet #1 contains four traffic words including 'street', 'sidewalk', 'left', and 'construction'. However, the tweet is asking a question but not reporting infromation about a transportation infrastructure. Analogously, the tweets #2 and #3 are reporting non-traffic events with their street level address although they consist of several traffic words. Such observation emerges the model weakness in classfying NT tweets with multiple traffic keywords. Unfortuentely, we have not observed a systematic incapbality for false-negative tweets.

# 6. Conclusions

Leveraging the power of word2vec models, we developed a robust and simple framework to detect traffic-related tweets on a real-time basis. The key idea for classifying a tweet is to measure the semantic similarity between words in the target tweet and a few traffic keywords. In addition to addressing the shortcomings in traditional BOW representation, our proposed model takes advantage of outstanding merits including: 1) Model is primarily contingent on unsupervised learning tools (i.e., word2vec models). 2) Model has only one trainable parameter (i.e., *Threshold*). 3) Model needs only a small set of traffic keywords (i.e., around 10), while it is insensitive to the number, source, and frequency of traffic keywords. 3) Model does not require any assumption to be made on distribution of training data and other models' variables. 4) Model does not need a large training set for neither extracting traffic keywords nor training the *Threshold* parameter. 5) Model is capable of detecting TR tweets that have no words in the selected traffic keywords. 6) Model takes all tweet's words into account for the classification task. Furthermore, the reported indexes such as test accuracy, precision, recall, and F-score reveal the superiority of the proposed model in attaining high-prediction quality. As an application, the proposed framework can be used by traffic management centers as a complementary source for real-time monitoring of traffic conditions.